\documentclass[twocolumn,superscriptaddress]{revtex4}
\usepackage[dvips]{graphicx}
\usepackage{float,color}

\begin{document}

\title{Signatures of the Helical Phase in the Critical Fields at Twin Boundaries of Non-Centrosymmetric Superconductors}

\author{Kazushi Aoyama}
\affiliation{The Hakubi Center for Advanced Research, Kyoto University, Kyoto 606-8501, Japan}
\affiliation{Department of Physics, Kyoto University, Kyoto 606-8502, Japan}
\author{Lucile Savary} 
\affiliation{Department of Physics, University of California, Santa Barbara, California 93106-9530, USA}
\author{Manfred Sigrist}
\affiliation{Institute for Theoretical Physics, ETH Zurich, Zurich 8093, Switzerland}

\begin{abstract}
Domains in non-centrosymmetric materials represent regions of different crystal structure and spin-orbit coupling. Twin boundaries separating such domains display unusual properties in non-centrosymmetric superconductors (NCS), where magneto-electric effects influence the local lower and upper critical magnetic fields. As a model system, we investigate NCS with tetragonal crystal structure and Rashba spin-orbit coupling (RSOC), and with twin boundaries parallel to their basal planes. There, we report that there are two types of such twin boundaries which separate domains of opposite RSOC. In a magnetic field parallel to the basal plane, magneto-electric coupling between the spin polarization and supercurrents induces an effective magnetic field at these twin boundaries. We show this leads to unusual effects in such superconductors, and in particular to the modification of the upper and lower critical fields, in ways that depend on the type of twin boundary, as analyzed in detail, both analytically and numerically. Experimental implications of these effects are discussed.  
\end{abstract}
\maketitle

\section{introduction}
Spin-orbit coupling is the cause of many extraordinary properties of materials, such as the anomalous and the spin Hall effects, topological insulators and superconductors \cite{AHE, SHE, TPI, TPS}. 
In the past decade, triggered by the discovery of the heavy Fermion superconductor CePt${}_3$Si  which lacks inversion symmetry \cite{CePt3Si}, studies of spin-orbit coupling effects on superconductivity have attracted much attention \cite{Springer}. Moreover, in the context of topological phases local properties of these non-centrosymmetric superconductors (NCS),  like the subgap states appearing at sample edges \cite{TPS, Vorontsov, Iniotakis_PRB} and domain boundaries \cite{Iniotakis,Arahata}, 
have been discussed. In our study, we address special properties of NCS with Rashba spin-orbit coupling (RSOC), which possess twin domains of opposite RSOC. In particular, we show that certain twin boundaries separating such domains can influence the superconducting (SC) properties of type-II superconductors in magnetic fields. 

The Rashba-type spin-orbit interaction \cite{Rashba} is inherent to systems lacking certain mirror symmetries. If $z \to -z $ is not a crystal symmetry then RSOC takes the basic form $\alpha  ({\bf k}\times \hat{z})\cdot {\bf S}$, with momentum ${\bf k}$, spin ${\bf S}$ and coupling constant $\alpha$. The NCS CePt$_3$Si \cite{CePt3Si}, and $f$- and $d$-electron NCS with the BaNiSn${}_3$-type crystal structure such as CeTSi$_3$ (T=Rh, Ir) \cite{CeRhSi3, CeIrSi3}, BaPtSi${}_3$ \cite{BaPtSi3}, and CaMSi$_3$ (M=Pt, Ir) \cite{CaIrSi3, CaPtSi3} belong to this class of Rashba-type superconductors.
One intriguing feature of Rashba-type NCS is the magneto-electric effect, which couples the spin polarization to supercurrents through spin-orbit coupling \cite{normal_jM, SC_jM, SC_Bj, Dimitrova,Samokhin, Kaur,Fujimoto}. A Zeeman field polarising electron spins thereby results in a spatial dependence of the phase of the SC order parameter following $ \Delta = \Delta_0 e^{i {\bf q}\cdot {\bf r}}$.  In this sense, this phase-modulated SC state is similar to a Fulde-Ferrell-Larkin-Ovchinnikov state \cite{FF, LO} and is known as helical SC phase \cite{Kaur}. The corresponding wave vector ${\bf q}\sim \alpha(\hat{z}\times \mu_B{\bf H})$ is oriented perpendicularly to both the magnetic field and the direction of the mirror symmetry-breaking (here the $z$-axis) if the electronic structure is nearly isotropic in $x$-$y$-direction. Despite the non-vanishing phase gradient there are no supercurrents flowing in the bulk of the system due to gauge invariance \cite{Kaur, Springer}. Therefore, the helical phase is generally difficult to detect. It has been proposed, however, that 
for inhomogeneous systems the helical phase could give rise to observable features. In two-dimensional NCS, such as the LaAlO${}_3$-SrTiO${}_3$ SC interface \cite{interface-SC,tunable-RSO}, where, for inplane fields, orbital depairing is suppressed, inhomogeneities can host magnetic flux patterns pointing perpendicular to the SC film and the applied field in the helical phase \cite{AS}. Also, in three-dimensional bulk materials, inhomogeneities can generate an unusual flux response to an external field via the helical phase, although in the latter case, vortices and orbital depairing effects could disturb the observation \cite{Kaur, Ikeda}. 

In our study, we address superconducting properties which are typical for certain twin boundaries in Rashba-type NCS with tetragonal crystal symmetry lacking the $z \to -z $  mirror symmetry,  like in CePt$_3$Si and the CeTSi$_3$ family. Twin domains in such materials have RSOC of opposite signs (in a sense we specify below). We consider here the case of domains which are stacked along the $z$-axis, separated by twin boundaries parallel to the basal plane of the crystal, as shown in Fig.~\ref{fig:twin}(a). For magnetic fields in the basal plane, the wave vector of the helical phase has opposite signs in the two twin domains, following the change of signs of the RSOC. The mismatch of the helical structures at the twin boundaries leads locally to supercurrents which cannot be screened completely, unlike in the bulk of the domains, as mentioned above. The resulting effective field influences the behavior of type-II superconductors in the mixed phase, i.e.\ between the lower and upper critical magnetic fields, $ H_{c1} $ and $ H_{c2} $, respectively. In particular, this magneto-electric effect actually affects the lower and upper critical fields, a phenomenon we address here. 
It is important to notice that, for domains stacked along the $z$-axis, there are two types of twin boundaries (see Fig.\ref{fig:twin}), which behave differently in a magnetic field. As we will find below the critical fields are shifted in opposite way at these two types of twin boundaries, in one case, being higher, and in the other, lower than the bulk value (see Fig.\ref{fig:twin}(b)).


The remainder of this paper is organized as follows. We first define the minimal model appropriate to eventually describe the novel features we report, and relevant to the bulk of a non-centrosymmetric superconductor with tetragonal symmetry. We then describe the different types of twin boundaries and the modifications we use to implement the existence of each type of twin-boundary.  In the following section we thoroughly investigate the upper critical field $H_{c2}$. There, we show that the effect of twin boundaries can be quite striking, and exhibit the different consequences of ``opposite'' types of twin boundaries. We then turn to the case of the lower critical field, and argue that the twin boundaries may act as pinning planes for vortices. In both cases, namely $H_{c1}$ and $H_{c2}$, we show both an analytical and a numerical analysis. Finally, we conclude and discuss experimental consequences.

\section{Model}

\begin{figure}[t]
\includegraphics[width=\columnwidth]{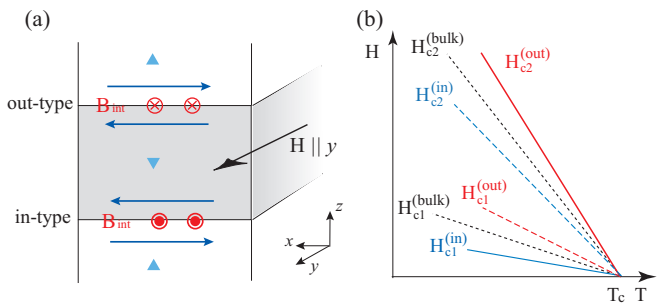}
\caption{(a) Crystal twin domains (white and gray regions) inside a single-crystal sample of a non-centrosymmetric superconductor, where the triangles denote the orientations of the axis of RSOC. The out (resp.\ in)-type twin boundary (parallel to the basal plane ($x$-$y$)) is described as a boundary with a positive (resp.\ negative) value of $\tilde{K}$ {(or equivalently $\delta N_0/N_0$)} in Eq.~(\ref{eq:twin}). An external magnetic field ${\bf H}$ applied parallel to the twin boundaries yields local internal fields ${\bf B}_{int}$ due to a mismatch of magneto-electric currents (blue arrows). (b) Schematic phase diagram of this system. Both upper and lower critical fields are shifted at the twin boundaries from their bulk values, suggesting that physical $H_{c2}(T)$ and $H_{c1}(T)$ curves (solid curves) are determined at the out- and in-type boundaries, respectively. \label{fig:twin}}
\end{figure}

Superconductivity in twinned materials has drawn much interest for a long time in part because the SC transition temperature can be enhanced at twin boundaries due to soft phonons along the boundary plane or distinct two-dimensional electronic states \cite{Buzdin_rev}. With such a $T_c$ enhancement, the upper and lower critical fields at twin boundaries should also locally be higher than the corresponding bulk values. In our study, we ignore the possibility of an enhanced SC critical temperature at the twin boundary, and assume a spatially uniform $T_c$. We focus, rather, on the influence of magneto-electric effects
in NCS in a magnetic field. The only feature of sample twinning which we take into account is the sign change of the coupling constant $ \alpha $ at the twin boundary. Moreover, we restrict the discussion to the case of a dominant $s$-wave SC channel and, in particular, for simplicity, we ignore odd-parity components which, on symmetry grounds, could be admixed \cite{Springer}.

The relevant Ginzburg-Landau (GL) theory can be derived from the BCS Hamiltonian including RSOC \cite{Kaur,Samokhin,Ikeda,AS}. 
The corresponding functional is obtained as usual as an expansion in the $s$-wave order parameter $ \Delta $,  
\begin{eqnarray} \label{eq:GL}
{\cal F}_{GL} & = & \int dz d{\bf r}_\perp \bigg[ a^{(2)} | \Delta|^2 + a^{(4)}  | \Delta|^4 + K_\perp |  {\bf \Pi}_\perp \Delta|^2 \nonumber\\
&+& K_z |  {\bf \Pi}_z \Delta|^2+ {K_{me}}(\hat{z} \times {\bf B} ) \cdot \{ \Delta^* {\bf \Pi}_\perp \Delta + c.c. \}  \nonumber\\
&+&  \frac{ (\nabla \times {\bf A})^2}{8 \pi} \bigg], 
\label{eq:GL}
\end{eqnarray}   
where the covariant gradient is defined as $ {\bf \Pi} = - i \hbar \nabla + (2e/c) {\bf A} $, where $ {\bf A}$ is 
the vector potential satisfying $\nabla \times{\bf A}={\bf B}$ with ${\bf B}$ the internal magnetic field, and where
\begin{equation}
a^{(2)} = N_0 \left( \ln \frac{T}{T_c} + 2 \gamma g^2 \mu_B^2 B^2 \right) 
\label{eq:a2}
\end{equation}
with $ T_c $ the bulk critical temperature, $\mu_B$ the Bohr magneton, $g$ the gyromagnetic ratio, $N_0=(N_++N_-)/2$ where $N_\pm$ denote the densities of states of the two bands split by the RSOC (see Appendix), and with $\gamma$, $K_\perp$, $K_z$, $K_{me}$ and $a^{(4)}$ given in the Appendix which also explains details of our notations.  The second term in $a^{(2)}$ (see Eq.~(\ref{eq:a2})) includes the paramagnetic pair-breaking effect through the Zeeman field $ g\mu_B {\bf B} $, and the last gradient term in Eq.~(\ref{eq:GL}) involving $K_{me}(\hat{z}\times {\bf B})$ introduces the magneto-electric effect which couples the spin polarization to the supercurrent.
This term changes signs under the mirror inversion $z \rightarrow -z$. Thus, we emphasise, it is only allowed in systems where $z\rightarrow-z$ is not a symmetry operation, and is therefore quite specific to NCS. 
Its coefficient, $K_{me}$, is connected to the RSOC and can be expressed as  
\begin{eqnarray}\label{eq:tilde-K}
K_{me} &=& \frac{\delta N_0}{N_0} g \mu_B \frac{K_\perp}{v_\perp} \nonumber\\
\frac{\delta N_0}{N_0} &=& \frac{N_+ - N_-}{(N_+ + N_-)/2} \propto \frac{\alpha}{E_F},
\end{eqnarray}
where $v_\perp$ is the in-plane Fermi velocity. Note that 
the sign of $ K_{me} $ is directly connected to the sign of the RSOC. 

For the following discussion, we introduce three characteristic length scales: the SC coherence length $\xi$, the magnetic length $r_H$, and the London penetration depth $\lambda_L$, defined as
\begin{eqnarray}\label{eq:length} 
\xi^{-2} &=& |a^{(2)}|/(\hbar^2K_\perp) \nonumber\\
r_H^{-2} &=& 2eH/(c\hbar) \nonumber\\
\lambda_{L}^{-2} &=& 32\pi K_\perp  |\Delta_0|^2e^2/c^2,
\end{eqnarray}
where $|\Delta_0|^2=|a^{(2)}|/(2a^{(4)})$, the uniform zero-field order parameter from the GL equations. 
For the in-plane field configuration, the bulk orbital-limiting 
and the paramagnetic-limiting (Pauli-limiting) fields at $T=0$ are given by
\begin{eqnarray}
H_{orb}(T=0) &=& \gamma_{\rm FS} \Phi_0/(2\pi \xi_0^2), \nonumber\\
H_p(T=0) &=& \pi T_c/(\sqrt{2}e^{\gamma_E}\mu_B g),
\end{eqnarray}
respectively, where {$\gamma_E\approx0.577$ is Euler's constant,} $\Phi_0=ch/(2e)$ is the magnetic flux quantum, $\xi_0=\hbar{v_\perp}/(2\pi T_c)$ is the in-plane SC coherence length at $T=0$, and {$\gamma_{\rm FS}=\sqrt{K_\perp/ K_z}$ parametrizes} the anisotropy of the Fermi surface. 
The strength of the Pauli-paramagnetic effect is quantified by the Maki parameter 
\begin{equation}
\alpha_M=\sqrt{2}H_{orb}(0)/H_p(0).
\end{equation}
In the following, for concreteness, we apply the magnetic field along the $ y $-axis and assume no spatial dependence along this direction. 

We turn now to a system with twin domains of 'up' ($\alpha > 0 $ or $ K_{me} > 0 $) and 'down' ($\alpha < 0 $ or $ K_{me} < 0 $) characters separated by twin boundaries with a geometry as shown in Fig. \ref{fig:twin}. The twin boundaries we consider are parallel to the $x$-$y$-plane. As mentioned in the introduction, we distinguish two types of twin boundaries, the `top-up bottom-down' (out-type) and `top-down bottom-up' (in-type) twin boundaries. It will become clear below that the two behave differently in a magnetic field parallel to the twin boundary plane. Within our GL model, only the sign of $K_{me}$ distinguishes the twin domains, as is reflected by $K_{me}\propto\alpha$ (see Eq.~(\ref{eq:tilde-K})). In practice, we implement the existence of twin boundaries by a sharp sign change of a space-dependent coefficient $ K_{me}(z) $:
\begin{equation}\label{eq:twin}
K_{me}(z)=\tilde{K} \, {\rm sgn}(z).
\end{equation} 
Because the change in the RSOC coefficient {$\alpha$} at the twin boundary happens on atomic length scales, the spatial variation of $K_{me} $ occurs on a much shorter length scale than the coherence length of the superconductor, so that the infinitely-abrupt change in $K_{me}$ implemented in Eq.~(\ref{eq:twin}) should therefore be qualitatively valid. 
Moreover, the existence of a sign change in {$K_{me}$} in Eq.~(\ref{eq:twin}) can be understood from the viewpoint of symmetry. If we take the twin boundary plane as a mirror reflection plane, the twin domain system is invariant under the corresponding mirror operation. Correspondingly, the magneto-electric term involving $K_{me}(z)(\hat{z} \times {\bf B})$ with the space dependent $K_{me}$ described by Eq.~(\ref{eq:twin}) does not change signs under this symmetry operation, leaving the free energy Eq.~(\ref{eq:GL}) invariant.

Throughout this paper, positive and negative values of $K_{me}$ will be assigned to crystal domains of `up' and `down' characters, respectively. Therefore, the out (resp.\ in)-type twin boundary in Fig.~\ref{fig:twin}(a) is described by positive (resp.\ negative) values of $\tilde{K}$ in Eq.~(\ref{eq:twin}).

\section{The upper critical field}

First we address the nucleation of superconductivity in high magnetic fields, in the presence of twin boundaries parallel to the basal plane. 
This can be discussed using the linearized GL equations with an unscreened external field ${\bf H}={\bf B}$ parallel to the twin boundary: 
the derivation of the instability condition of the normal state, which yields the upper critical field $ H_{c2} $ necessitates no more. Therefore we need only consider the terms quadratic in $\Delta$ in Eq.~(\ref{eq:GL}). This quadratic form will be denoted ${\cal F}_{\rm GL}^{(2)}$ in what follows. 

We choose the gauge such that the vector potential is ${\bf A}=z H\hat{x}$ for a field along the $y$-direction, and we impose periodic boundary conditions along the $x$-direction. This allows us to represent the order parameter as
\begin{equation}
\Delta({\bf r}) = \sum_{n} C_{n}(z) \,  e^{i \frac{2 \pi n}{L_x} x}
\end{equation}
with $ L_x $ the linear extension of the system in the $x$-direction.
First we tackle the problem variationally to obtain insight into the role of the twin boundary on $ H_{c2} $. The validity of the variational approach will be confirmed later by the comparison to a numerical solution of the linearized GL equation. 

\subsection{Variational approximation}

The standard way to determine the upper critical field is equivalent to finding and solving the ground state of the Schr\"odinger equation for a one-dimensional harmonic oscillator introduced by the vector potential $ {\bf A} (z) $. For our gauge choice, this harmonic potential confines the order parameter along the $z$-axis with its center at the twin boundary. However, here, the potential is modified through the additional $ K_{me}(z) $ term in $ {\cal F}_{\rm GL}^{(2)}$ which effectively introduces a small shift of the center in opposite directions on either side of the twin boundary. Still, at large distances away from the twin boundary the potential looks essentially harmonic and the following variational ansatz for the order parameter is therefore justified, 
\begin{equation}
C_n(z)=C_n \frac{1}{\sqrt{l_n\sqrt{\pi}}}e^{-z^2/2l_n^2},
\label{eq:Cn}
\end{equation}
where the length scales $l_n$ are variational parameters which will be determined so as to minimize the free energy, ${\cal F}^{(2)}_{\rm GL}$. 
Inserting Eq.~({\ref{eq:Cn}) into ${\cal F}_{\rm GL}^{(2)}$, we obtain
\begin{eqnarray}\label{eq:GL_2_0}
&&{\cal F}^{(2)}_{\rm GL} = \sum_n\frac{|C_n|^2}{l_n\sqrt{\pi}}\int_{-\infty}^\infty dz \; e^{-z^2/l_n^2} \bigg[ \;  a^{(2)} 
\nonumber \\
&& + \frac{K_z\hbar^2}{l_n^4}z^2  +  K_\perp P_n(z)^2 - 2 \, \tilde{K} {\rm sgn}(z)P_n(z) H \bigg] \nonumber \\
\end{eqnarray}
with
\begin{equation}
P_n(z) =   \frac{2\pi n \hbar}{L_x} +\frac{2eH}{c}z .
\label{pn}
\end{equation}
In the absence of the twin boundary, $K_{me}(z)$ is just a constant $K_{me}(z)=\tilde{K}$. 
Then, as we will see in Eq.~(\ref{eq:GLeq_Hc2}), the last term in Eq.~(\ref{eq:GL_2_0}) only yields an overall shift of the center of the harmonic potential and therefore has no effect on the orbital depairing field. (We will also find --see the right-hand side of Eq.~(\ref{eq:GLeq_Hc2})-- that the paramagnetic depairing is suppressed by $\tilde{K}$ \cite{Kaur}.) 
With the twin boundary, however, we encounter a real deformation of the potential. We can evaluate the integral Eq.~(\ref{eq:GL_2_0}) analytically, 
\begin{eqnarray}
{\cal F}^{(2)}_{\rm GL} &=& \sum_n |C_n|^2 \bigg[ a^{(2)} + K_\perp \hbar^2 \Big(\frac{2\pi n}{L_x}\Big)^2 +  \frac{K_\perp}{2} \Big(\frac{2eH}{c} \Big)^2 l_n^2 \nonumber\\
&& + \frac{K_z\hbar^2}{2}\frac{1}{l_n^2}-\frac{2}{\sqrt{\pi}}\frac{2eH}{c}\tilde{K}H l_n \bigg].
\end{eqnarray}
The different Fourier components $ C_n $ remain decoupled and we see immediately that only $ n =0 $ minimizes the variational free energy, resulting in 
\begin{eqnarray}
\frac{{\cal F}^{(2)}_{\rm GL}}{|C_0|^2} &=& a^{(2)} + \frac{\gamma_{\rm FS} K_z \hbar^2}{r_H^2} f_{\tilde{l}_0} \nonumber\\
 f_{\tilde{l}_0}&=& \frac{1}{2}\bigg[  \tilde{l_0}^2 + \frac{1}{\tilde{l_0}^2}-\frac{2}{\sqrt{\pi}}\frac{2\tilde{K}H r_H}{\sqrt{\gamma_{\rm FS}K_\perp K_z \hbar^2}} \, \tilde{l_0} \bigg],
\end{eqnarray}
where $\tilde{l_0}=l_0 \gamma_{\rm FS}^{1/2}/r_H$. 
For fixed values of the field $H$, we minimize $f_{\tilde{l}_0}$ with respect to $\tilde{l}_0$, and then, the SC transition point (the highest transition temperature) is  determined by the condition $a^{(2)} + (\gamma_{\rm FS} K_z \hbar^2/r_H^2) f_{\tilde{l_0},{\rm min}}=0$, i.e.,
\begin{equation}\label{eq:Tc_var}
\ln\frac{T}{T_c} = -\gamma T_c^2 \Big[f_{\tilde{l_0},{\rm min}}\frac{2\pi^2 H}{H_{orb}(0)} + \Big(\frac{\pi \alpha_M }{\sqrt{2}e^{\gamma_E}}\frac{H}{H_{orb}(0)}\Big)^2 \Big],
\end{equation}
where $f_{\tilde{l_0},{\rm min}}$ is the minimum value of the function $f_{\tilde{l_0}}$. 
The contribution of the magneto-electric effect is incorporated in 
\begin{eqnarray}
&&\frac{2\tilde{K}H r_H}{\sqrt{\gamma_{\rm FS}K_\perp K_z\hbar^2}} = \frac{1}{2 e^{\gamma_E}} \frac{\delta N_0}{N_0} \alpha_M \sqrt{\frac{H}{H_{orb}(0)}}.
\end{eqnarray}

\begin{figure}[t]
\includegraphics[scale=1.1]{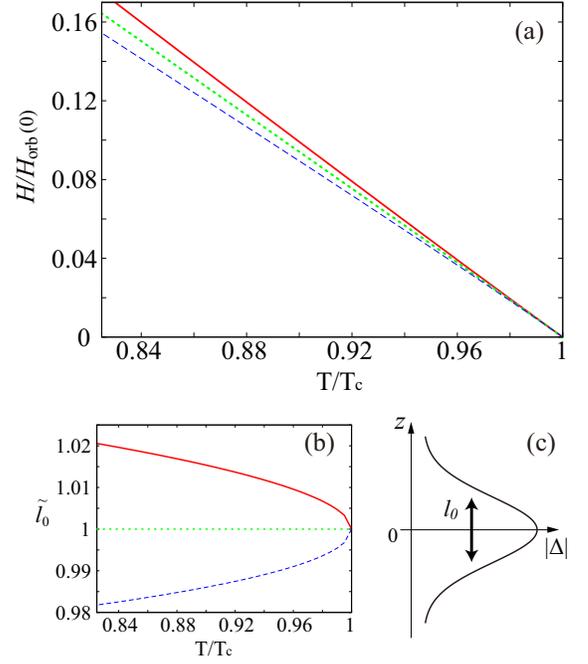}
\caption{(color online) The SC instability in the bulk (green dotted curves) and at the twin boundaries with $\delta N_0/N_0=0.4$ (red solid lines) and $\delta N_0/N_0=-0.4$ (blue dashed lines) for Maki parameter $\alpha_M=3$. (a) Temperature dependence of the upper critical field $H_{c2}(T)$ and (b) the corresponding behavior of the {\it effective} magnetic length $\tilde{l}_0$, which, as depicted in (c), measures the extent of the SC pairing function along the $z$ axis centered at the twin boundary. This length scale $\tilde{l}_0$ is normalized by its bulk value. \label{fig:Hc2_M3}}
\end{figure}

Now we address the two types of twin boundaries, distinguished here by the sign of $ \tilde{K} $, corresponding to the out ($ \tilde{K} > 0$)- or in-type ($ \tilde{K} < 0$) twin boundary as shown in Fig.\ref{fig:twin}.  
Figure \ref{fig:Hc2_M3} (a) displays $H_{c2}(T)$ curves for the nucleation of the superconducting order parameter at the twin boundary with a moderate paramagnetic effect. For positive values of $\tilde{K}$ (out-type), the upper critical field at the twin boundary is enhanced compared to the bulk value, while for negative values of $\tilde{K}$ (in-type), it is lower than the bulk $H_{c2}$. In the latter case, superconductivity would surely appear first in the bulk and would be rather suppressed at the twin boundary. To understand why $H_{c2}(T)$ is enhanced or suppressed at the twin boundaries, we examine the {\it effective} magnetic length $l_0$.

Fig.~\ref{fig:Hc2_M3} (b) shows the temperature dependence of $\tilde{l}_0$ (for which the free energy is minimized), which measures the extent of the order parameter along the $z$-axis. For positive $\tilde{K}$, the effective magnetic length $l_0=\tilde{l}_0 \gamma_{FS}^{1/2} r_H$ is larger than the corresponding bulk value, so that $ \Delta (z) $ is more extended. This can be interpreted in terms of an {\it effective} magnetic field $H_{eff}$ at the twin boundary, lower than the applied field:  $ H_{eff} = H / \tilde{l}_0^2 $. In contrast, for negative $ \tilde{K} $, the effective field is enhanced at the twin boundary, suppressing the nucleation of SC there. This is consistent with the picture that the mismatch of the helical modulations in the two adjacent domains is compensated by an internal field which is added to or subtracted from the external field. Note that this magneto-electric effect depends on the Zeeman coupling and the stronger the paramagnetic limiting effect, the more pronounced it is. In Fig. \ref{fig:Hc2_M8}  we show $H_{c2}(T)$ curves for a stronger paramagnetic effect, i.e.\ with a larger Maki parameter $\alpha_M$. There, besides the relative enhancement of the shift of the local $ H_{c2} $, we also observe that the temperature dependence is different from the basically linear increase below $ T_c $ in Fig.\ref{fig:Hc2_M3}. The rather strongly bent curve of $ H_{c2} $ seen here originates from the dominant paramagnetic-limiting compared to the orbital-limiting regime \cite{Adachi, Mineev_para,CeCoIn5_kappa}.

\begin{figure}[t]
\includegraphics[scale=0.55]{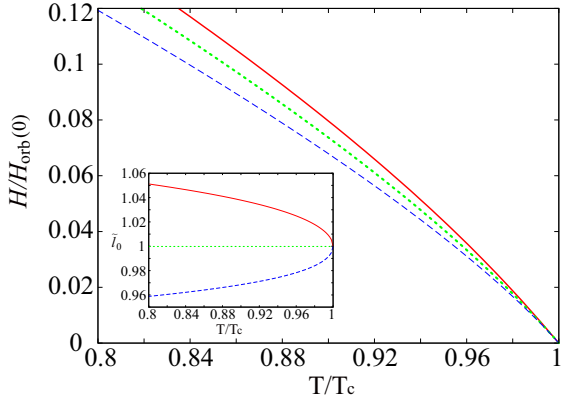}
\caption{Temperature dependence of the upper critical field $H_{c2}(T)$ and the {\it effective} magnetic length $\tilde{l}_0$ (inset) for large Maki parameter $\alpha_M=8$, with the same notations as in Fig.\ref{fig:Hc2_M3}. \label{fig:Hc2_M8}}
\end{figure}

\subsection{Numerical solution of the GL equation}

Now we turn to the numerical evaluation of the linearized GL equations, which allows us to assess the validity of our variational approach. 
We determine $ C_n(z) $ from the differential equation obtained by variationally differentiating ${\cal F}_{\rm GL}^{(2)}$ with respect to the order parameter,
\begin{eqnarray}\label{eq:GLeq_Hc2}
&&\left[ \partial_{\tilde{z}}^2 - \Big(\tilde{z} + \frac{\tilde{K} {\rm sgn}(\tilde{z}) H r_H}{\sqrt{\gamma_{\rm FS} K_\perp K_z \hbar^2}} - \frac{2\pi n}{L_x} r_H \gamma_{\rm FS}^{1/2}\Big)^2 \right] C_n(\tilde{z}) \nonumber\\
&&  \qquad = \frac{r_H^2}{\gamma_{\rm FS}}\Big( \frac{a^{(2)}}{K_z \hbar^2}-\frac{\tilde{K}^2H^2}{K_\perp K_z \hbar^2}\Big) C_n(\tilde{z}) 
\end{eqnarray}
with $\tilde{z}=z\gamma_{\rm FS}^{1/2}/r_H$ a dimensionless coordinate. Because the solution of interest is symmetric under $ z \to - z $, we choose $ n=0 $. This eigenvalue equation is most-efficiently solved by expanding $ C_0(\tilde{z}) $ in the basis of wave functions of the harmonic oscillator
\begin{eqnarray}
C_0(\tilde{z}) &=& \sum_m u_m \, \varphi_m(\tilde{z}),\nonumber\\
\varphi_m(\tilde{z}) &=& \frac{e^{-\tilde{z}^2/2}}{\sqrt{2^m m! \sqrt{\pi}}} H_m(\tilde{z}),
\end{eqnarray}
where $H_m(\tilde{z})$ are the Hermite polynomials. Since $\varphi_m(\tilde{z})$ satisfies the eigenvalue equation
\begin{equation}
\big(\partial_{\tilde{z}}^2 -  \tilde{z}^2 \big) \, \varphi_m(\tilde{z}) = -(2m+1) \varphi_m(\tilde{z}),
\end{equation} 
the GL equation can be rewritten as,
\begin{eqnarray}
&& \sum_m M_{lm}  u_m = -\frac{r_H^2 a^{(2)}}{\gamma_{\rm FS}K_z \hbar^2}   u_l , \nonumber\\
&& M_{lm} =  (2m + 1)\, \delta_{l,m} -  \frac{2\tilde{K}H r_H}{\sqrt{\gamma_{\rm FS}K_\perp K_z \hbar^2}}V_{lm}, \nonumber\\
&& V_{lm} =  [1+(-1)^{l+m}] \int_0^\infty d\tilde{z} \, \tilde{z} \varphi_l(\tilde{z})\varphi_m(\tilde{z}),
\end{eqnarray}
where the relation $H_m(-\tilde{z})=(-1)^m H_m(\tilde{z})$ has been used. Note that $ V_{lm} $ is symmetric, $V_{lm}=V_{ml}$. 
The problem is reduced to finding the eigenvalues of the matrix $M_{lm}$. The superconducting instability follows from the equation $-(r_H^2 a^{(2)})/(\gamma_{\rm FS}K_z \hbar^2)=\lambda_{\rm min}$, such that
\begin{equation}\label{eq:Tc_num}
\ln\frac{T}{T_c} = -\gamma T_c^2 \Big[\lambda_{\rm min}\frac{2\pi^2 H}{H_{orb}(0)} + \Big(\frac{\pi \alpha_M }{\sqrt{2}e^{\gamma_E}}\frac{H}{H_{orb}(0)}\Big)^2  \Big],
\end{equation}
where $\lambda_{\rm min}$ is the minimal eigenvalue of $M_{lm}$.
At this point, we notice that $\lambda_{\rm min}$ in Eq.~(\ref{eq:Tc_num}) corresponds to $f_{\tilde{l_0},{\rm min}}$ in Eq.~(\ref{eq:Tc_var}), so that the validity of the variational approach can be checked by comparing $\lambda_{\rm min}$ and $f_{\tilde{l_0},{\rm min}}$. As one can see in  Fig.\ref{fig:comp_M8}, the two values $\lambda_{\rm min}$ and $f_{\tilde{l_0},{\rm min}}$ coincide well at all temperatures, suggesting that our variational approach is a good approximation and also validating the interpretation. 

\begin{figure}[t]
\includegraphics[width=\columnwidth]{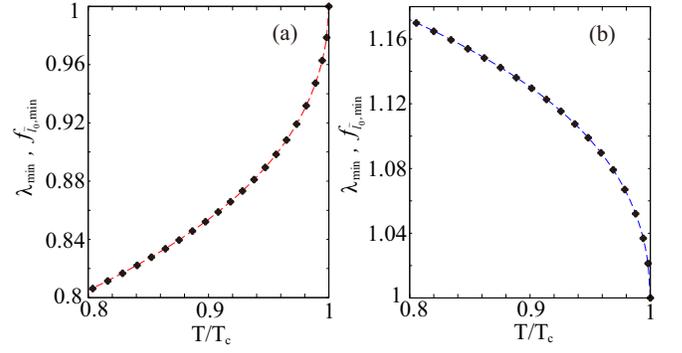}
\caption{Comparison between the result obtained by the variational method $f_{\tilde{l}_0,{\rm min}}$ (dashed curves) and the corresponding numerical result $\lambda_{\rm min}$ (circles) for $\delta N_0/N_0=0.4$ (a) and $\delta N_0/N_0 =-0.4$ (b). The same Maki parameter as Fig. \ref{fig:Hc2_M8}, $\alpha_M=8$, is used. \label{fig:comp_M8}}
\end{figure}

\section{The lower critical field}

In this section we address the effect of twin boundaries on the lower critical field. For this purpose we investigate the line energy of a single vortex on
the twin boundary. Contrary to the previous section, we consider first the numerical solution, and then turn to a variational discussion in the London limit to give some insight into the mechanism. In order to simplify the discussion, and because we expect the results to not be qualitatively affected by this restriction, we assume an isotropic situation by setting $K_z=K_\perp$. This allows us to formulate the problem simply in cylindrical coordinates  $(x,y,z)=(r\cos\theta,y,-r\sin\theta)$ with the magnetic field pointing, again, along the $ y $-axis. 

\subsection{ Magnetic flux distribution and $H_{c1}(T)$} 

For the following discussion it will be convenient to express the order parameter and the vector potential in their Fourier expansion with respect to $\theta$,
\begin{eqnarray}
\Delta(x,z) &=& |\Delta_0|\sum_n c_n(r) e^{in\theta} , \nonumber\\
A_{\theta}(r,\theta) &=& \frac{c\hbar}{2e\xi}\sum_m a_m(r) e^{im\theta}.
\end{eqnarray}
Here, $A_{\theta}(r,\theta)$ is related to the vector potential in the Cartesian coordinate system through the equation $A_x(x,z)\hat{x}+A_z(x,z)\hat{z} = A_\theta(r,\theta) \hat{\theta}$, and both $c_n(r)$ and $a_n(r)$ are assumed to take real values only.
By substituting these expressions into ${\cal F}_{\rm GL}$, Eq.~(\ref{eq:GL}), and carrying out the integral with respect to $\theta$, we obtain the GL free energy density per unit length in the $y$ direction, defined through,
\begin{eqnarray}\label{eq:GL_cylindrical}
{\cal F}_{\rm GL}&=& 2\pi |\Delta_0|^2K_\perp \hbar^2 \int_0^\infty \tilde{r}d\tilde{r} f_{\rm GL}, 
\end{eqnarray}
with
\begin{eqnarray}
f_{\rm GL} &=&  \sum_n \Big( -c_n^2 + [\partial_{\tilde{r}} c_n]^2+\frac{n^2}{\tilde{r}^2}c_n^2\Big) \nonumber\\
&+& \frac{1}{2}\sum_{n_1,n_2,n_3,n_4}c_{n_1}c_{n_2}c_{n_3}c_{n_4}\delta_{n_1+n_2,n_3+n_4} \nonumber\\
&+& \sum_{n,n',m} c_n c_{n'} a_m \Big(\frac{2n}{\tilde{r}}\delta_{n,n'+m} + \sum_{m'} a_{m'} \delta_{n+m,n'+m'} \Big)\nonumber\\
&+& \frac{c\tilde{K}}{4eK_\perp\xi}\sum_{n,n',m'} \Big(\frac{1}{\tilde{r}}\partial_{\tilde{r}}\big[\tilde{r}a_{m'}\big]\Big) \Big( D^{(1)}_{n,n',m'} c_{n'}(\partial_{\tilde{r}}c_n) \nonumber\\
&& -  D^{(2)}_{n,n',m'}\frac{n+n'}{2\tilde{r}}c_{n'}c_n -\sum_m D^{(3)}_{n,n',m,m'} c_n c_{n'}a_m  \Big) \nonumber\\
&+& \frac{\lambda_L^2}{\xi^2}\sum_m \Big(\frac{1}{\tilde{r}}\partial_{\tilde{r}} [\tilde{r} a_m]\Big)\Big(\frac{1}{\tilde{r}}\partial_{\tilde{r}} [\tilde{r} a_{-m}]\Big).
\end{eqnarray}
Here, $\tilde{r}=r/\xi$ and
\begin{eqnarray}
&& D^{(1)}_{n,n',m'} = i\big[d_{n+m'-n'}^{(+)} - d_{n'+m'-n}^{(+)}\big], \nonumber\\
&& D^{(2)}_{n,n',m'} = i\big[d_{n+m'-n'}^{(-)} + d_{n'+m'-n}^{(-)}\big], \nonumber\\
&& D^{(3)}_{n,n',m,m'} = i \big[d_{n+m+m'-n'}^{(-)} + d_{n-m+m'-n'}^{(-)} \big], \nonumber
\end{eqnarray}
with 
\begin{eqnarray}
&& d_n^{(\pm)} = \int_0^{2\pi} d\theta \frac{\tilde{K}(z)}{2\pi \tilde{K}}(e^{i(n+1)\theta}\pm e^{i(n-1)\theta}) \nonumber\\ \\
&& = \left\{ \begin{array}{ll}
\delta_{n,-1} \pm \delta_{n,1} & \mbox{bulk} \\ & \\
\displaystyle \frac{2}{i \pi} \big[\frac{1}{n+1}\pm\frac{1}{n-1}\big]\delta_{n,even} & \mbox{twin boundary},
\end{array}\right. 
\end{eqnarray}
where the upper (resp.\ lower) case is for a vortex far from (resp.\ right on) the twin boundary. 
The magnetic field is given by
\begin{equation}
B(r,\theta)=\frac{1}{r}\partial_r \big[r A_{\theta}(r,\theta) \big]=\frac{c\hbar}{2e\xi}\frac{1}{r}\partial_r \big[r \sum_m a_m(r)e^{im\theta} \big],
\end{equation} 
and $a_m(r)=a_{-m}(r)$ holds. Note that, therefore, in the bulk without twin boundaries, the magneto-electric term proportional to $\tilde{K}$ vanishes and 
does not affect the line energy of the vortex.

Now, since a single vortex with its singularity at $r=0$ contains the total flux $\Phi_0$, we have the limiting conditions, for one vortex centered at $r=0$,
\begin{equation}
\Delta (r,\theta) = |\Delta_0|e^{i\theta}  
\label{eq:27}
\end{equation}
for $ r \rightarrow \infty $ and $ \Delta (r =0 ,\theta) = 0 $ as well as,
\begin{eqnarray}
\Phi_0 &=& \int_0^{2\pi}d\theta\int_0^\infty r dr B(r,\theta) \nonumber\\
&=& 2\pi\frac{c\hbar}{2e\xi}\big[ \lim_{r \rightarrow \infty}r a_0(r)-\lim_{r \rightarrow 0}r a_0(r) \big].
\end{eqnarray}
Note that, because the magnetic field vanishes far from the vortex core ($B(r,\theta)\rightarrow 0 $ for  $r\rightarrow\infty$), the magneto-electric term proportional to $B(r,\theta)$ is not active at large distances from the vortex center, and thus, there, the condition for a {\em usual} single vortex $\Delta (r,\theta) = |\Delta_0|e^{i\theta} $, Eq.~(\ref{eq:27}), can be used even in the case with the twin boundary. 
Now, the above constraints lead to the boundary conditions on $c_n(r)$ and $a_m(r)$, 
\begin{eqnarray}\label{eq:bd-condition}
&& \left.\begin{array}{l}
c_n (\tilde{r}) = \delta_{n,1} \\
{\tilde{r}} a_m(\tilde{r})= -\delta_{m,0} 
\end{array}\right\} \,  \mbox{for} \; \tilde{r} \rightarrow \infty  , \nonumber\\ 
&& \left. \begin{array}{l}
c_n (\tilde{r}) = 0 \\
{\tilde{r}} a_0(\tilde{r})= -2 \\
\partial_{\tilde{r}} \big[{\tilde{r}} a_{m \neq 0}(\tilde{r}) \big]= 0
\end{array}\right\} \, \mbox{for} \; \tilde{r} \rightarrow 0 .
\end{eqnarray}

The single vortex energy per unit length along the vortex axis is given by
\begin{equation}
e_v = 2\pi   \int_0^\infty r dr \, \bigg[ \frac{|\Delta_0|^2 K_\perp \hbar^2}{\xi^2} f_{\rm GL}-\Big(-\frac{ |a^{(2)}|^2}{4 a^{(4)}} \Big)\bigg],
\end{equation}
and leads to 
the lower critical field, 
\begin{eqnarray}
\frac{H_{c1}(T)}{H_{orb}(0)}&=& \frac{4\pi}{\Phi_0}e_v \Big/\frac{\Phi_0}{2\pi \xi_0^2}\nonumber\\
&=&   \Big(\frac{\xi_0}{\xi}\Big)^2 \Big(\frac{\xi}{\lambda_L}\Big)^2 \frac{1}{2}\int_0^\infty \tilde{r} d\tilde{r} \, \Big(f_{\rm GL}+\frac{1}{2}\Big).
\end{eqnarray}
Here, the Zeeman term in $a^{(2)}$ has been dropped because it is negligibly small at low fields, near $H_{c1}$, for any reasonable value of the Maki parameter.


\begin{figure}[t]
\includegraphics[scale=0.58]{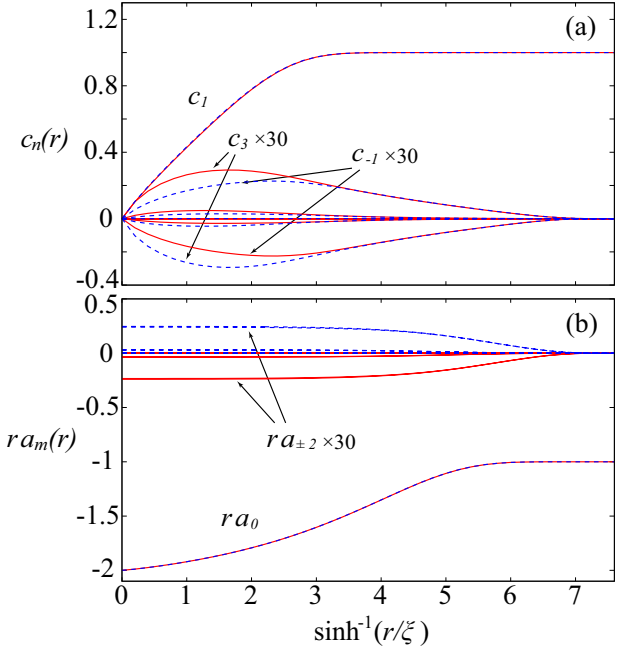}
\caption{Radial dependences of $c_n(r)$ (a) and $ra_m(r)$ (b) for the twin boundaries with $\delta N_0/N_0=0.4$ (red solid curves) and $\delta N_0/N_0=-0.4$ (blue dashed ones) at $ T/T_c =0.85$, where the parameters $\alpha_M=8$ and $\lambda_L/\xi=10$ are used. Without twin boundaries, only $c_1$ and $a_0$ are nonvanishing with almost the same spatial dependences as displayed here. All the components except $c_1$ and $a_0$ are multiplied by 30. \label{fig:ca_dep}}
\end{figure}

By numerically solving the GL equations $\delta {\cal F}_{\rm GL}/\delta c_k=0$ and $\delta {\cal F}_{\rm GL}/\delta a_k=0$ under the constraints of Eq.~(\ref{eq:bd-condition}), we investigate the spatial structure of $c_k(\tilde{r})$ and $a_k(\tilde{r})$.  As a typical example, in Fig.\ref{fig:ca_dep} we plot spatial profiles of $c_k(\tilde{r})$ and $a_k(\tilde{r})$ for the large Maki parameter $\alpha_M=8$. One can see that, in contrast to the bulk case, where only $c_1$ and $a_0$ are nonvanishing, additional components $c_{1\pm2}$ and $a_{\pm2}$ appear near the vortex center induced by the twin boundary. Since $a_{\pm 2}$ involves the phase factor $e^{\pm i2\theta}$, finite values of these components suggest the occurrence of a deformation of the magnetic flux distribution on the twin boundary. 
Also note that the sign of $a_{\pm 2}$ depends on the sign of $\tilde{K}$.  

Figure \ref{fig:Hc1} (a) shows the $H_{c1}(T)$ curves at the two twin boundaries and in the bulk. The effect of the twin boundary on the temperature dependence of $H_{c1}$ is qualitatively the same as that for $H_{c2}$: the lower critical field is enhanced (suppressed) for positive (negative) values of $\tilde{K}$. The $\tilde{K}$-dependent behavior of $H_{c1}$ is natural because, as we have discussed in the previous section, positive $\tilde{K}$ yields a counter vortex field, while negative $\tilde{K}$ effectively strengthens the magnetic field stabilizing the vortex. This effect of the twin boundary can be also seen in the magnetic flux distribution. We introduce two length scales measuring the extension of the flux distribution in the $x$- and $z$-directions, $W_x$ and $W_z$, which are defined by 
\begin{equation}
W_{i}=\int_0^\infty dr r B(r,\theta_i) \Big/ \int_0^\infty dr B(r,\theta_i)
\end{equation}
with $\theta_x=0$ and $\theta_z=\pi/2$.

\begin{figure}[t]
\includegraphics[scale=0.58]{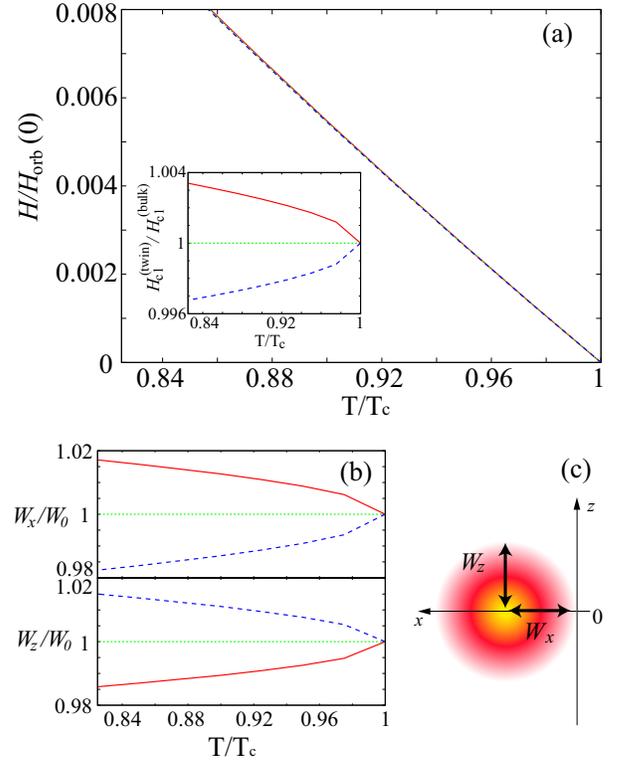}
\caption{The single-vortex instability in the bulk (green dotted curves) and at the twin boundaries with $\delta N_0/N_0=0.4$ (red solid lines) and $\delta N_0/N_0=-0.4$ (blue dashed lines) for $\alpha_M=8$. (a) Temperature dependence of the lower critical field $H_{c1}(T)$ and (b) the corresponding behavior of the spatial extent of the magnetic flux along the $x$-axis $W_x$ (upper panel) and along the $z$-axis $W_z$ (lower panel). The magnetic flux distribution of a single-vortex with its core located at the twin boundary is sketched in (c). The inset of (a) shows the ratio of $H_{c1}(T)$ at the twin boundary to its bulk value. \label{fig:Hc1}}
\end{figure}

Figure \ref{fig:Hc1} (b) shows the temperature dependence of $W_x$ and $W_z$ normalized by the bulk value $W_0$. In the bulk, $W_x=W_z$ is satisfied because we assumed isotropy. For positive $\tilde{K}$, the magnetic flux is extended in the $x$-direction and squeezed in the $z$-direction, leaving the total flux to be $\Phi_0$. This anisotropy is caused by the magnetic field induced through the magneto-electric coupling. 
For positive $\tilde{K}$ the effective field on the twin boundary is smaller than the bare field of the vortex, so that the stability of superconductivity against the bare field is higher on the twin boundary than away from it. Thus, the magnetic flux extends along the twin boundary ($x$-direction) to lower the energy. Conversely, for negative $ \tilde{K} $ the induced field is opposite, leading to a flux distribution compressed along the $x$-direction. 

\subsection{Extended London model} 

We will now focus on the line energy of a vortex on a twin boundary using an extended London theory incorporating the magneto-electric coupling. 
For this purpose we fix the shape of the vortex in the London limit as $\Delta(x,z)=\theta(r-\xi)|\Delta_0|e^{i\phi(x,z)}$ with the radius $r=\sqrt{x^2+z^2}$
and the step function $\theta(r)$ taking care of the fact that the vortex core extends over a coherence length $ \xi $, and $\phi$ a smooth real  function of space coordinates. 
In this limit, the magnetic field ${\bf B}$, the SC current ${\bf j}$, and the vortex-line energy $e_{v0}$ for an ordinary $s$-wave superconductor are given by
\begin{eqnarray}\label{eq:London}
&& {\bf B}(x,z) = \hat{y} \, \frac{\Phi_0}{2\pi\lambda_L^{2}} K_0 \Big(\frac{r}{\lambda_L} \Big), \nonumber\\
&& {\bf j}= -4eK_\perp|\Delta_0|^2\Big( \hbar  \nabla \phi + \frac{2e}{c} {\bf A} \Big)=\frac{c}{4\pi}\big(\nabla \times {\bf B} \big), \nonumber\\
&& e_{v0} \simeq \Big(\frac{\Phi_0}{4\pi \lambda_L} \Big)^2 \ln \Big( \frac{\lambda_L}{\xi}\Big),
\end{eqnarray}
where $K_0(x)$ is a  modified Bessel function \cite{Tinkham}. 
Using the expression of Eq.~(\ref{eq:London}), we evaluate variationally the change of the vortex-line energy $\delta e_v$ due to the magneto-electric coupling by simply adding the integral of $K_{me}(\hat{z}\times {\bf B}) \cdot {\bf j} $ in Eq.~(\ref{eq:GL}), which leads to
\begin{eqnarray}
\delta e_v &=& -\frac{1}{2eK_\perp} \int dxdz \, K_{me}(z) B(x,z) \, j_x(x,z) \nonumber\\ \\
&=& \left\{ \begin{array}{ll}
0 & \mbox{bulk} \\ & \\ \displaystyle
\frac{\pi c \tilde{K}}{4 e K_\perp \xi } \frac{\xi}{\lambda_L} \left(\frac{\Phi_0}{4\pi \lambda_L}\right)^2 & \mbox{twin boundary}.
\end{array}\right. .
\end{eqnarray}
The total vortex energy in the presence of the twin boundaries, $e_v=e_{v0}+ \delta e_v $, is then
\begin{equation}\label{eq:variation}
\frac{e_v}{e_{v0}} \simeq 1 + R_v \,  \frac{\delta N_0}{N_0} \frac{\alpha_M \sqrt{1-T/T_c}}{\big(\lambda_L/\xi \big) \, \ln\big(\lambda_L/\xi \big)} ,
\end{equation}
where $R_v=\pi \, [e^{\gamma_E}2\sqrt{14\zeta(3)} ]^{-1}=0.215$. 
Equation (\ref{eq:variation}) shows good agreement with the numerical result shown in the inset of Fig.\ref{fig:Hc1} (a), with the $\sqrt{|T-T_c|}$ dependence, as well as with the rather small difference $\delta e_v \propto H_{c1}^{\rm (twin)}-H_{c1}^{\rm (bulk)}$.   
The shift of $H_{c1}(T)$ due to the twin boundaries increases with increasing RSOC, i.e. with increasing $\delta N_0/N_0 \propto \alpha $, and with increasing Pauli-paramagnetic effect quantified by the Maki parameter $\alpha_M$, but is diminishes with increasing GL parameter $\kappa = \lambda_L/\xi$. 

We may also view $ \delta e_v $ as the potential energy of a vortex, which is zero in the bulk, but varies smoothly as the twin boundary is approached. This potential is repulsive for positive $ \tilde{K} $ and attractive for negative $ \tilde{K} $. In the latter case vortices can more easily penetrate the sample along the twin boundary than into the bulk. Thus, vortices should line up on this type of twin boundary. Conversely, when $ \tilde{K} $ is positive, vortices avoid twin boundaries, which are then (weak) barriers for the crossing of vortices. Quantitatively, however, this local shift of the lower critical field is much weaker than that of the upper critical field and is most likely not of experimental relevance. 


\section{Conclusion}

We have examined the influence of magneto-electric effects on the upper and lower critical fields in a non-centrosymmetric superconductor with twin boundaries. Considering the case of tetragonal crystal symmetry with Rashba spin-orbit coupling, appropriate for example for twin boundaries in CePt$_3$Si, we found that two types of twin boundaries parallel to the basal plane exist, which separate domains of opposite RSOC. Magneto-electric effects which are irrelevant for the behavior in the bulk, enhance or reduce the upper and lower critical fields at the twin boundaries depending on the type of the latter. Although our analysis is based on a Ginzburg-Landau formulation for an $s$-wave order parameter and ignores the admixture of an odd-parity pairing component, the results obtained should be qualitatively valid beyond the temperature range where the GL theory presented here is valid. 

We found that the effect on the lower critical field is most likely too small to be observed, but the fact that for one type of twin boundary the upper critical field
is enhanced could indeed be of experimental relevance. Since the volume fraction of the crystal that is actually influenced by the twin boundaries is generally small, experimental probes quite sensitive to superconductivity such as magnetic torque and AC susceptibility should provide the best tools to detect the enhanced $H_{c2}$ at the twin boundary. As we have seen, some twin boundaries suppress $H_{c2}$ and $ H_{c1} $ and, thus, may act as pinning planes for vortices in the mixed phase. Any non-centrosymmetric material, as discussed in our model, should display alternating in- and out-type twin boundaries, such that both kinds of observable features, i.e., the enhanced $ H_{c2} $ (out-type), as well as the vortex pinning (in-type) due to a reduced $ H_{c2} $, could be potentially seen in such a single sample. 
Together with the observation of these features, detecting crystal domains directly with a real space imaging method would also provide important information to investigate further novel effects due to twin boundaries, as addressed here.

Finally, we would like to note that one may create a twin-boundary-like structure by contacting two crystals of opposite RSOC to one another along the $z$-axis, forming a planar Josephson junction. In that case also, two types of Josephson junctions exist, and, in particular, the Josephson vortices do display distinct features \cite{Savary}. 

We are grateful to G. Eguchi, S. Yonezawa and Y. Maeno for motivating discussions. This work is supported by a Grant-in-Aid for Scientific
Research (Grant No. 25800194) and a grant by the Swiss Nationalfonds. Moreover, K.A. thanks to the Pauli Centre for Theoretical Studies of ETH Zurich for hospitality during his stay.

\section{Appendix}

\subsection{GL coefficients in Eq.~(\ref{eq:GL})}
The GL coefficients in Eq.~(\ref{eq:GL}) have been derived elsewhere \cite{Kaur,Samokhin,Ikeda,AS} and are given by
\begin{eqnarray}
a^{(4)} &=& N_0 \gamma , \nonumber\\
K_\perp &= &N_0 \gamma \langle v_{x,y}^2 \rangle , \quad K_z = N_0 \gamma \langle v_{z}^2 \rangle , \nonumber\\
K_{me} & =&  \delta N_0 \,  g \mu_B \, \gamma \, v_\perp/2 , \nonumber\\
\gamma &=&  \frac{7 \zeta(3)}{16 (\pi k_B T_c)^2} , \nonumber\\ 
 N_0 &=& (N_+ + N_-)/2 , \quad  \delta N_0 = N_+ - N_- ,
\end{eqnarray}
where $ N_{\pm} $ denote the density of states of the two bands split by the RSOC, $v_i$ denotes the Fermi velocity in the $i$ direction, $\langle A \rangle $ represents the angle average of $A$ on the Fermi surface, and $v_\perp=\sqrt{\langle v_x^2 + v_y^2  \rangle}$.
In deriving Eq.~(\ref{eq:GL}), we restrict ourselves to $|\alpha|/E_F \ll 1$ and $g\mu_BH/|\alpha| \ll 1$.

{ \subsection{GL equations for Eq.~(\ref{eq:GL_cylindrical})} }
The saddle point equations with respect to $c_k$ and $a_k$, $\delta {\cal F}_{\rm GL}/\delta c_k=0 $ and $\delta {\cal F}_{\rm GL}/\delta a_k=0$, yield the GL equations
\begin{widetext}
\begin{eqnarray}
 \Big(\partial^2_{\tilde{r}}+\frac{1}{\tilde{r}}\partial_{\tilde{r}}- \frac{k^2}{\tilde{r}^2}\Big)c_k &=& -c_k+\sum_{n_1,n_2,n_3}c_{n_1}c_{n_2}c_{n_3}\delta_{n_1+n_2,n_3+k} +\sum_{n,m,m'} c_n a_m a_{m'} \delta_{n+m,k+m'} +\sum_{n}\frac{c_n}{\tilde{r}}\Big( n a_{n-k} +k a_{k-n} \Big)\nonumber\\
&& +\frac{c\tilde{K}}{8eK_\perp\xi}\sum_{n,m'}  \bigg[ -\frac{1}{\tilde{r}}D^{(1)}_{k,n,m'} \, c_n \partial^2_{\tilde{r}}\big[\tilde{r}a_{m'}\big]  + \frac{1}{\tilde{r}}\partial_{\tilde{r}}\big[\tilde{r}a_{m'}\big] \Big( \big[ D^{(1)}_{n,k,m'} - D^{(1)}_{k,n,m'} \big](\partial_{\tilde{r}}c_n) \nonumber\\
&& \qquad \qquad \qquad - \big[D^{(2)}_{n,k,m'}+D^{(2)}_{k,n,m'} \big]\frac{n+k}{2\tilde{r}}c_n  -\sum_m \big[D^{(3)}_{n,k,m,m'} + D^{(3)}_{k,n,m,m'} \big]c_na_m  \Big) \bigg]
\end{eqnarray}
and
\begin{eqnarray}
 \frac{\lambda_L^2}{\xi^2}\Big(\partial^2_{\tilde{r}}+\frac{1}{\tilde{r}}\partial_{\tilde{r}}- \frac{1}{\tilde{r}^2}\Big)a_{-k} &=& \sum_{n,n',m}c_n c_{n'}a_m\delta_{n+k,n'+m} +\sum_{n}\frac{n+k}{\tilde{r}}c_n c_{n+k} - \frac{c\tilde{K}}{8eK_\perp\xi}\sum_{n,n'} \bigg[ \sum_m D^{(3)}_{n,n',k,m} c_n c_{n'}\frac{1}{\tilde{r}}\partial_{\tilde{r}}\big[\tilde{r}a_{m}\big] \nonumber\\
 && \quad +\partial_{\tilde{r}} \Big( D^{(1)}_{n,n',k} c_{n'}(\partial_{\tilde{r}}c_n)- D^{(2)}_{n,n',k}\frac{n+n'}{2\tilde{r}}c_n c_{n'} -\sum_m D^{(3)}_{n,n',m,k} c_n c_{n'}a_m  \Big) \bigg],
\end{eqnarray}
\end{widetext}
respectively.

\end{document}